\documentclass{elsart}
\usepackage{latexsym}
\usepackage{epsfig}
\usepackage{natbib}

\def\url#1{#1} 

\begin{document}

\begin{frontmatter}

\title{A review of the properties of the scattered starlight which 
contaminates the spectrum of reddened stars}

\author{Fr\'ed\'eric \snm Zagury}
\address{
\cty 02210 Saint R\'emy Blanzy, \cny France
\thanksref{email} }

   \thanks[email]{E-mail: fzagury@wanadoo.fr}

\accepted{February 2002}

 \begin{abstract}

     When a star is observed behind an interstellar cloud of 
sufficient column density, we do not observe the direct light from the 
star, which is totally extinguished.
Rather, we see only starlight scattered at small angles from the star.

I use several papers published in New Astronomy to 
recount the different steps which permit understanding how, and under 
which conditions, scattered 
starlight can be more important than direct starlight in the spectrum 
of a reddened star.
Associated problems -the fit of the extinction curve, the nature of the scatterers 
(small grains, or, atoms or molecules from the gas), the $2200\,\rm\AA$ bump-  
are also, briefly, discussed.
\end{abstract} 
\end{frontmatter}
 \section{Introduction} \label{intro}
In a preceding paper \citep{uv7} I have reviewed several arguments which prove 
that the standard interpretation of the extinction curve, upon which 
all interstellar dust models are based, does not agree with observation.
If the standard theory is wrong, it must be that the spectrum of 
reddened stars comprises in the UV an important part of starlight scattered at 
small angular distance from the stars \citep{uv7}.

We already know that all the intensity of this light is 
concentrated within $0.3"$ from the star \citep{uv7}.
I will use results from previously published papers to pinpoint the nature 
of this light and of the scatterers.
This forecasts new directions of research which are discussed in 
section~\ref{cons}.
\section{The spectrum of the scattered light} \label{spectre}
\subsection{Importance of the scattered light} \label{int}
The importance of the scattered light was roughly estimated in 
\citet{uv2}.
In the UV spectrum of stars with an intermediate reddening ($E(B-V)\sim 0.3$) 
we can separate the direct starlight, which still dominates the near-UV 
part of the spectrum, and the scattered starlight, clearly superimposed on 
the tail of the exponential decrease of the direct starlight.
The scattered starlight can reach $\sim 15-20\%$ of the direct starlight corrected for extinction \citep{uv2}.
As a comparison, the light we receive from the star is extinguished by 
$\sim e^{-2E(B-V)/\lambda}$ ($\lambda$ in $\mu$m), if the linear visible 
extinction extends to the UV.
For $1/\lambda \sim 6\mu$m$^{-1}$, and $E(B-V)=0.3$, the direct 
starlight is less than $3\%$ of what it would be if there was no 
extinction.
It is under $3/1000$ for $E(B-V)=0.5$.
We see that if the reddening is sufficient, the scattered starlight largely dominates the direct starlight.
\subsection{Wavelength dependence of the scattered light} \label{forme}
Since scattering occurs very close from the star, scattered and 
direct starlights must cross the same medium, thus being extinguished 
in a similar manner.
By correcting the spectrum of a reddened star normalized by the 
spectrum of a non-reddened star of same spectral type (the reduced 
spectrum of the star, \citet{uv2}) we expect to find a constant in the 
visible (1 if the non-reddened star and the reddened one are identical 
and at the same distance from the observer), and a spectrum which 
caracterizes the scattered light in the UV.

\begin{figure*}
\resizebox{!}{0.8\columnwidth}{\includegraphics{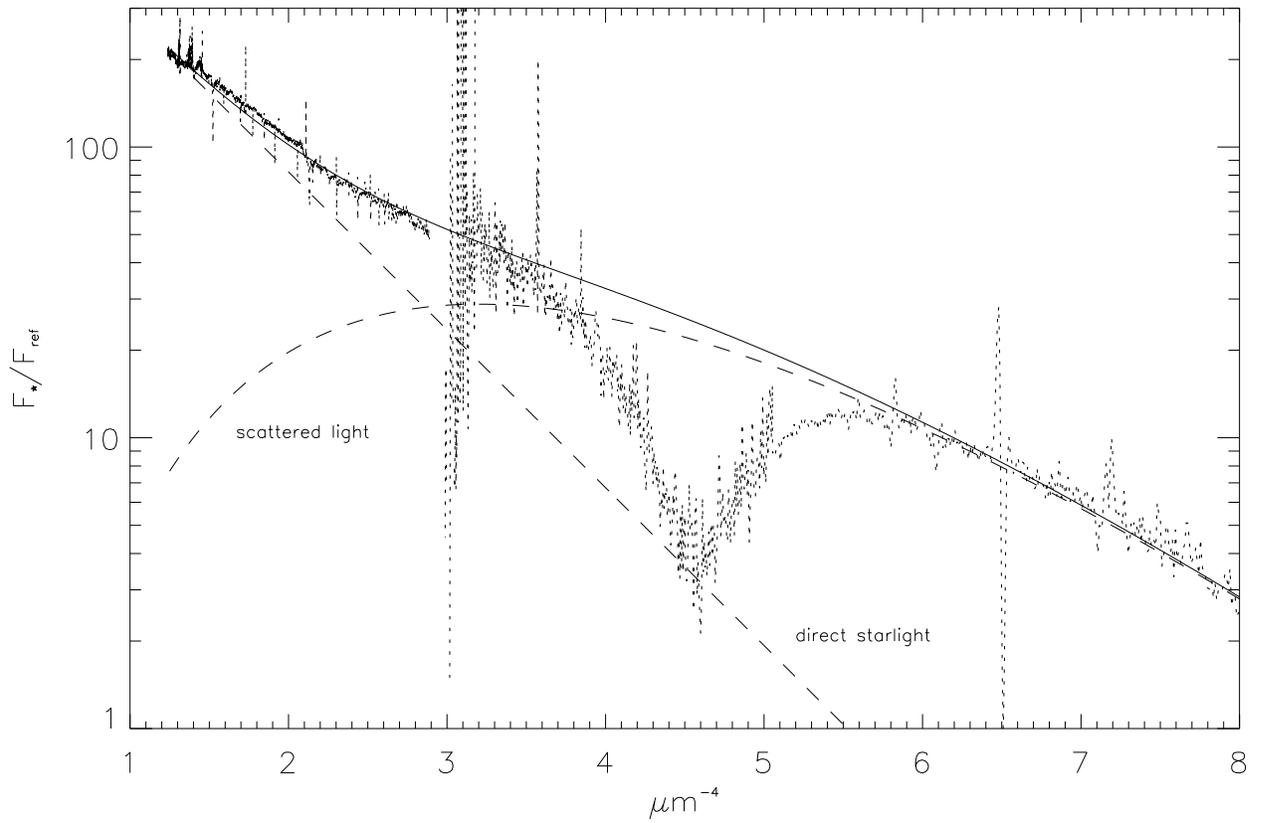}} 
\caption{The reduced spectrum of HD46223 and its' decomposition, 
outside the bump region into direct starlight and scattered light. The 
plain curve is the sum of the two components.
} 
\label{fig:hd46223}
\end{figure*}

The method was applied \citep{uv3} to the reduced spectrum of HD46223 
($E(B-V)\sim 0.5$) which was established from the visible to the far-UV.
It was finally found that,
outside the bump region, the reduced spectrum of HD46223 is the sum 
of two components (see figure~\ref{fig:hd46223}, adapted from \citet{uv3}).
One is the extinction of the direct starlight, proportional to 
$e^{-2E(B-V)/\lambda}$. 
The second is the scattered light, $\propto \lambda^{-4}e^{-2E(B-V)/\lambda}$.
Hence, if we correct the reduced spectrum of HD46223 for extinction 
we find that the scattered light depends on wavelength as $\lambda^{-4}$.
\section{The relation between visible and far-UV extinctions} \label{vis-far-UV}
If the scattered starlight depends on wavelength as $\lambda^{-4}$, 
the product of the reduced spectrum of a reddened star by $\lambda^{4}$ 
will be proportional to $e^{-2E(B-V)/\lambda}$ in the far-UV.

For a number of reddened stars, I have compared \citep{uv6} the value 
$E_{uv}$ of $E(B-V)$ found for the slope of the product (reduced 
spectrum)x$\lambda^{4}$ in the far-UV, to the value $E_{vis}$ given by 
the spectral type of the stars and $B-V$.
The agreement between $E_{uv}$ and $E_{vis}$ is excellent, with a 
tendency for $E_{uv}$ to be larger than $E_{vis}$.
This tendency is expected since scattered light, when it appears in 
the visible, diminishes the slope of the visible extinction.

This general relation between visible and UV extinctions is 
incompatible with the standard theory of extinction. 
According to which, there is no relation between the proportion of large grains responsible for the visible 
extinction, and the proportion of molecules responsible for the far-UV 
rise of the extinction curve \citep{jenniskens93}.
\section{Scattering at nul angle and scattering by the nebulae} 
\label{2grains}
In the UV, the scattering of starlight by a nebula follows a 
$1/\lambda$ law \citep{uv1, uv7}.
The particles responsible for the brightness of the nebulae are not
the same as the particles responsible for the additive component of scattered 
light in the spectrum of reddened stars, which scatter starlight as $1/\lambda^{4}$.

Light scattering as $1/\lambda$ is produced by large 
grains which preferentially scatter light forward.
A $1/\lambda^{4}$ scattering cross section (Rayleigh scattering) 
implies the presence of small particles, with a quasi-isotropic phase function.
Why is the near-isotropic scattering by small particles detected in the forward 
direction only, and not in the nebulae, while the forward scattering 
by the large grains is observed at larger angles but is 
negligible in the spectrum of reddened stars?
\section{Coherent scattering at very small angle} 
\label{coherent}
There is one possible answer to the preceding questions, and it is 
the only one I have found which satisfies all the 
requirements imposed by  observation.

While scattered light is at the first order proportional to the number 
of scatterers, it becomes proportional to the square of this number if 
scattering is coherent and if the phase lag between two scattered 
waves received by the observer is small.
If the particles are randomly distributed, this  
happens in the complete forward direction only \citep{uv4, bohren}.

Coherent scattering will then occur if the particles are identical, if 
they can consider the source-star as a point source \citep{berkeley}, 
and if the length difference between the star-observer distance and 
the path one scattered photon follows is small compared to the 
wavelength.

The two latter conditions have the following mathematic expression:
\begin{eqnarray}
     \theta&\ll & 3\,10^{-12}"\left(\frac{\lambda}{2000\,\rm\AA}\right)
     \left(10\frac{\Phi_{S}}{\Phi}\right)
     \left(\frac{d_{0}}{l_{0}}\right)
    \label{eq:c1}  \\
    \theta & \ll &  
    5\,10^{-8}"\left(\frac{\lambda}{2000\,\rm\AA}\right)^{0.5}
     \left(\frac{100\,pc}{l_{0}}\right)^{0.5}
     \left(\frac{d_{0}}{D}\right)^{0.5}
    \label{eq:c2}
\end{eqnarray}¥
$l_{0}$ is the observer-cloud distance, $d_{0}$ the star-cloud 
distance, $D$ the star-observer distance ($D=d_{0}+l_{0}$), $\phi$ and 
$\phi_{S}=1.4\,10^{6}\rm km$ the diameters of the star and of the sun, 
$\theta$ the angular distance viewed from the observer within which 
scattering is coherent.

In \citet{uv2} I made a mistake. I have neglected 
condition~\ref{eq:c1} which in most cases is the constraining one.
We can estimate a typical order of magnitude of $12"$ for the 
angular distance from the star within which scattered starlight is 
coherent.
For a cloud at $100\,\rm pc$ this represents $\sim 10^{6}\,\rm cm^{2}$ 
($\sim 100\,\rm m^{2}$).
If $\beta$ is the proportion of scatterers relative to the number of 
hydrogen atoms, for a typical $N_{H}=10^{20}\,\rm H/cm^{2}$ column 
density, coherent scattered light is $\beta 10^{26}$ times what it 
would be if scattering was incoherent.  
\section{Consequences and prospectives} \label{cons}
\subsection{Nature of the scatterers} \label{sca}
Thierry Lehner, from the Observatoire de Meudon, suggested to me that the 
scatterers could be atoms or molecules from the gas.
This hypothesis is attractive both because the gas is the 
principal component of interstellar clouds and because it ensures that 
the scatterers are identical.
According to \citet{rayleigh}, a set of atoms or molecules will 
diffuse as $\lambda^{-4}e^{\alpha\lambda^{-4}}$ ($\alpha$ a constant 
proportional to the number of particles).
This expression will be consistent with the observations if 
$\alpha\lambda^{-4}$ is negligible in the UV.
Diffusion by atomic or molecular hydrogen, because of the 
symetry of the particle, may introduce 
a $\lambda^{-6}$ term (in addition to the $\lambda^{-4}$ one) 
in the scatterring cross section (see \citet{jackson}).

According to \citet{sellgren} small grains are necessary to explain 
the infrared spectrum of interstellar clouds. 
These grains, if they exist, can also do the coherent scattering.
\subsection{Column density of scatterers} \label{colsca}
To estimate the column density of scatterers (in the cases of 
hydrogen and of small grains in the following) we need the scattering 
cross section $\sigma$, which depends on wavelength.

A $15\%$ ratio between scattered light and direct starlight corrected for 
extinction, corresponds to a colunm density $N$ of scatterers:
\begin{equation}
 N \sim 1.2\frac{d_0}{l_{0}D} 
 \sigma^{-0.5}\theta^{-2}
    \label{eq:n1}
\end{equation}¥
Relation~\ref{eq:c1} then implies:
\begin{equation}
 N \gg  1.7\,10^{13}\frac{l_0}{D} 
 \frac{100\,\mathrm{pc}}{d_0}
 \left(\frac{\Phi}{10\Phi_S}\right)^{2}¥
 \sigma_{0}¥^{-0.5}
    \label{eq:n2}
\end{equation}¥
with $\sigma_{0}=\sigma(\lambda/2000\,\rm\AA)^{2}$.

The scattering cross section of hydrogen for Rayleigh scattering is 
calculated from the polarisability: 
$\sigma_{0}=2.3\,10^{-4}\rm cm^{2}$.

For small spherical grains: 
$\sigma_{0}=3.2\,10^{-8}(a/100\,\rm nm)^{6}\rm cm^{2}$ 
(\citet{vandehulst}, section~6.4).
$a$ is the size of the particles.

Thus, respectively for hydrogen and for small grains:
\begin{eqnarray}
    N_{H}& \gg & 10^{27}\,\mathrm{cm}^{-2}\frac{l_0}{D} 
 \frac{100\,\mathrm{pc}}{d_0}
 \left(\frac{\Phi}{10\Phi_S}\right)^{2}
    \label{eq:nh}  \\
    N_{grains}& \gg & 10^{17}\,\mathrm{cm}^{-2} 
 \frac{100\,\mathrm{pc}}{d_0}
 \left(\frac{\Phi}{10\Phi_S}\right)^{2}
\left(\frac{100\,\mathrm{nm}}{a}\right)^3
    \label{eq:ng}
\end{eqnarray}¥
These calculations are given for sake of completness.
I do not have the practice of diffusion experiments which is necessary 
to judge of their validity.
Should they be correct, coherent scattering by hydrogen is to be 
ruled out.
\subsection{Analytic fit of the extinction curve} \label{fit}
{\small \emph{`Donnez-moi quatre param\`{e}tres et je vous dessine un \'{e}l\'{e}¥phant; 
donnez m'en un cinqui\`{e}me et il aura une trompe.'
citation attribu\'{e}e \`{a} Joseph Bertrand, math\'{e}maticien fran\c{c}ais (1822-1900)
}}

The traditional fit of the extinction curve by
\citet{fitzpatrick} requires five to six parameters (three for the bump region).
\citet{fitzpatrick} decomposition is a purely mathematical approximation 
with no physical meaning at all.
Except for maybe the bump region, if the bump truly is absorption by some 
specific type of grain.
The validity domain of the Fitzpatrick~\& Massa fit is 
restricted to the UV: I have checked on some examples that the extension of the 
fit to the visible often diverges from what is observed.
What predictive value can we give to a fit which is limited to 
the UV domain it was made for?
Is it a surprise if, using this fit, no relation was found between the three parts the 
standard theory distinguishes in the extinction curve, between the 
visible and the far-UV in particular?

The separation of the extinction curve in direct starlight and scattered 
light, the fit of each of these parts by the corresponding mathematic 
expression, the relation between visible and far-UV extinctions 
(section~\ref{vis-far-UV}), prove that it is possible to give a physical 
meaning to the fit of the extinction curve, and to
diminish the number of parameters necessary to fit the curve \citep{uv3}.
I believe a study of the variations of the 
extinction curve (in the visible and in the UV) for a large sample of stars observed behind one 
cloud will bring a better understanding of the extinction curve and 
will permit to fix its' exact number of free parameters.
\subsection{The $2200\,\rm\AA$ bump} \label{bump}
The comparative study of different observations recalled in the first 
sections of this paper casts new light on the near-UV part and the far-UV rise of the 
extinction curve.
The usual explanation of the bump, as a classical extinction process, can apply 
to this new context.
But, as it is mentionned in \citet{uv7}, it is not obvious that this is the case.

Both in the fit found for HD46223 (figure~\ref{fig:hd46223}), and 
in Seaton's curve (figure~1 in \citet{uv7}),
the extinction of the direct starlight approximately 
passes by the extremum of the bump.
Thus, the possibility for the bump to be an interruption of the scattered 
light only, and not of the direct starlight, can not be neglected.

Here again observations of many stars behind a cloud may be the way 
to understand how the bump is created.
\section{Conclusion} \label{conc}
The standard interpretation of the extinction curve was established 
from a phenomenological and straightforward reading of its' different 
parts.
The efforts developped to give a physical support to 
this interpretation led to sophisticated, but never satisfying, interstellar dust models.
In \citet{uv7} I have used arguments of common sense to prove that this 
rushed reading of the extinction curve is not in agreement with the observation.
I present that the only alternative to the standard interpretation, 
namely that the spectrum of a reddened star is contaminated by 
scattered light, logically solves the main problems the 
extinction curve poses: its' interpretation, the relation between 
its' mathematical representation and physics.

The light we receive from a reddened star is contaminated by 
scattered light. 
The intensity of the scattered starlight is important if the scattered waves are coherently 
added.
The angular distance from a reddened star, as seen by the observer, within which 
scattering is coherent, is small, of order $10^{-12}"$ from 
the star.
For sufficient reddening the far-UV scattered starlight is much larger than 
the direct starlight: in this case the far-UV light we receive from 
the direction of the star is nearly all scattered 
light, the star is not directly observed (it is totally extinguished).

The exact extinction curve of starlight by dust grains is a straight 
line from the near-infrared to the far-UV.

The comparison between the UV scattering law in the nebulae and the 
law found for the scattered light which contaminates the spectrum of 
reddened stars shows that two kind of particles are necessary  to 
explain the extinction of starlight.
Large grains, with a forward scattering phase function and an 
extinction cross-section $\propto 1/\lambda$;
and small particles (atoms, molecules or dust), which diffuse according to Rayleigh's 
law, with a near-isotropic phase function and a scattering cross-section $\propto 
1/\lambda^{4}$.

In this context, there is no more reason to 
suppose there are variations of the average properties of interstellar dust from one 
region to another.

Several questions are still open.
The most important ones are the 
nature of the  $2200\rm\AA$ bump, the exact number of degrees of freedom
the extinction curve has, and the nature of the scatterers.
The $2200\rm\AA$ bump may not be a classical 
extinction process by some kind of dust grain.
Only the scattered starlight seems to be extinguished in the 
bump region, leaving the direct starlight unaffected.
The degrees of freedom of the extinction curve (6 in the 
\citet{fitzpatrick} decomposition) are already diminished.
I believe that a study of the extinction curve, if it can be 
established for a large set of stars, observed behind the same cloud, 
in the optical and in the UV, will permit the understanding of the origin of 
the bump, and constrain the number of free parameters in the 
extinction curve.
{}
\end{document}